**A facile approach to calculating superconducting transition temperatures in the bismuth solid phases.**

Isaías Rodríguez[1], David Hinojosa-Romero[2], Alexander Valladares[1], Renela M. Valladares[1] and Ariel A. Valladares[2] *

[1] Facultad de Ciencias, Universidad Nacional Autónoma de México, Apartado Postal 70-542, Ciudad Universitaria, CDMX, 04510, México

[2] Instituto de Investigaciones en Materiales, Universidad Nacional Autónoma de México, Apartado Postal 70-360, Ciudad Universitaria, CDMX, 04510, México.

* Corresponding Author: Ariel A. Valladares, valladar@unam.mx

**All solid phases of bismuth under pressure, but one, have been experimentally found to superconduct. From Bi-I to Bi-V, avoiding Bi-IV, they become superconductors and perhaps Bi-IV may also become superconductive. To investigate the influence of the electronic properties $N(E)$ and the vibrational properties $F(\omega)$ on their superconductivity we have *ab initio* calculated them for the corresponding experimental crystalline structures, and using a BCS approach have been able to determine their critical temperatures $T_c$ obtaining results close to experiment: For Bi-I (The Wyckoff Phase) we predicted a transition temperature of less than 1.3 mK and a year later a $T_c$ of 0.5 mK was measured; for Bi-II $T_c$ is 3.9 K measured and 3.6 K calculated; Bi-III has a measured $T_c$ of 7 K and 6.5 K calculated for the structure reported by Chen *et al.*, and for Bi-V $T_c$ ~ 8 K measured and 6.8 K calculated. Bi-IV has not been found to be a superconductor, but we have recently predicted a $T_c$ of 4.25 K.**

1. Introduction

When superconductivity was discovered by Kamerlingh Onnes in mercury in 1911 ($T_c$ ~ 4.2 K) [1] after he produced liquid helium for the first time in 1908 (boiling point of 4.2 K at atmospheric pressure) this phenomenon was assumed only to consist of a vanishing electrical resistance for some metallic materials. It was not until the Meissner effect (the expulsion of magnetic fields) was observed that this discovery became a puzzle. Now superconductivity is displayed by many and sundry materials and is characterized by exhibiting zero electrical resistance and the above-mentioned Meissner effect below a characteristic temperature $T_c$, the superconducting transition temperature.

In 1957 Bardeen, Cooper and Schrieffer (BCS) [2] developed the first successful theory of superconductivity based on two simple but revolutionary and decisive concepts. They proposed that electrons pair through the atomic vibrations in the material due to the now known Cooper pairing potential, and that this pairing gives rise to the transition to the superconducting state; also, that the coherent motion of the paired electrons gives them the inertia to sustain electrical currents without dissipation [2]. Alternative ideas have appeared since then and even different concepts to substitute the initial ones but BCS has withstood the passing of time. Since phonons are invoked to be responsible for the electron pairing, a manifestation of this interaction should appear in superconductivity, and it does: the so-called isotope effect which is the dependence of $T_c$ on the isotopic mass, $T_c \sqrt{M}$ = const. The Meissner effect is also borne out; hence the two main aspects of superconductivity are duly accounted for by the BCS theory.

Due to the variety and abundance of materials that superconduct, it has also been ventured that in principle all materials may become superconductors if cooled down to low enough temperatures. In what follows we demonstrate that invoking the corresponding electronic densities of states, $N(E)$, and the vibrational densities of states, $F(\omega)$, for the various solid phases of bismuth under pressure, the superconducting transition temperatures can be calculated if the Cooper attraction sets in with a strength comparable for all phases. This facile approach, if proven correct, can be generalized to study phases of other materials similarly related.

BCS was a breakthrough that fostered a remarkable growth of the field propitiating progress and new questionings and generating alternative theories to explain specific phenomena. However, in 2016 based on the original BCS we predicted that at ambient pressure bismuth becomes a superconductor at a temperature below 1.3 mK [3]. A year later an experimental group corroborated that in fact Bi is a superconductor with a transition temperature of 0.5 mK, a result that eluded previous work [4]. Analogously we have studied low dimensional structures of bismuth, bilayers or bismuthene, and have predicted that they may become superconductive, with a transition temperature of $T_c{}^L$ = 2.61 K, provided the Cooper pairing potential manifests itself in a similar manner as in the bulk, [5, 6]. Finally, since it is very suggestive that all phases of bismuth under pressure should superconduct we calculated $N(E)$ and $F(\omega)$ for Bi-IV and predicted a transition temperature of 4.25 K [7], to be corroborated.

Why does Bi display several crystalline phases at moderate pressures? Five solid phases, from Bi-I to Bi-V, have been identified for pressures below 7 GPa and for temperatures between 400 and 600 K [7, 8]. It seems that all crystalline structures are energetically accessible in this interval of the P-T plane, and that differences among them respond more to subtle changes in the binding energy per atom; mass densities are also relatively similar, from 9.8 g/cm$^3$ for Bi-I to 12.64 g/cm$^3$ for Bi-V. Fig. 1 displays an accepted present-day classification of the bismuth phases where an increasing tendency for the experimental $T_c$ exists as the pressure is increased, at least up to 7 GPa [7, 9].

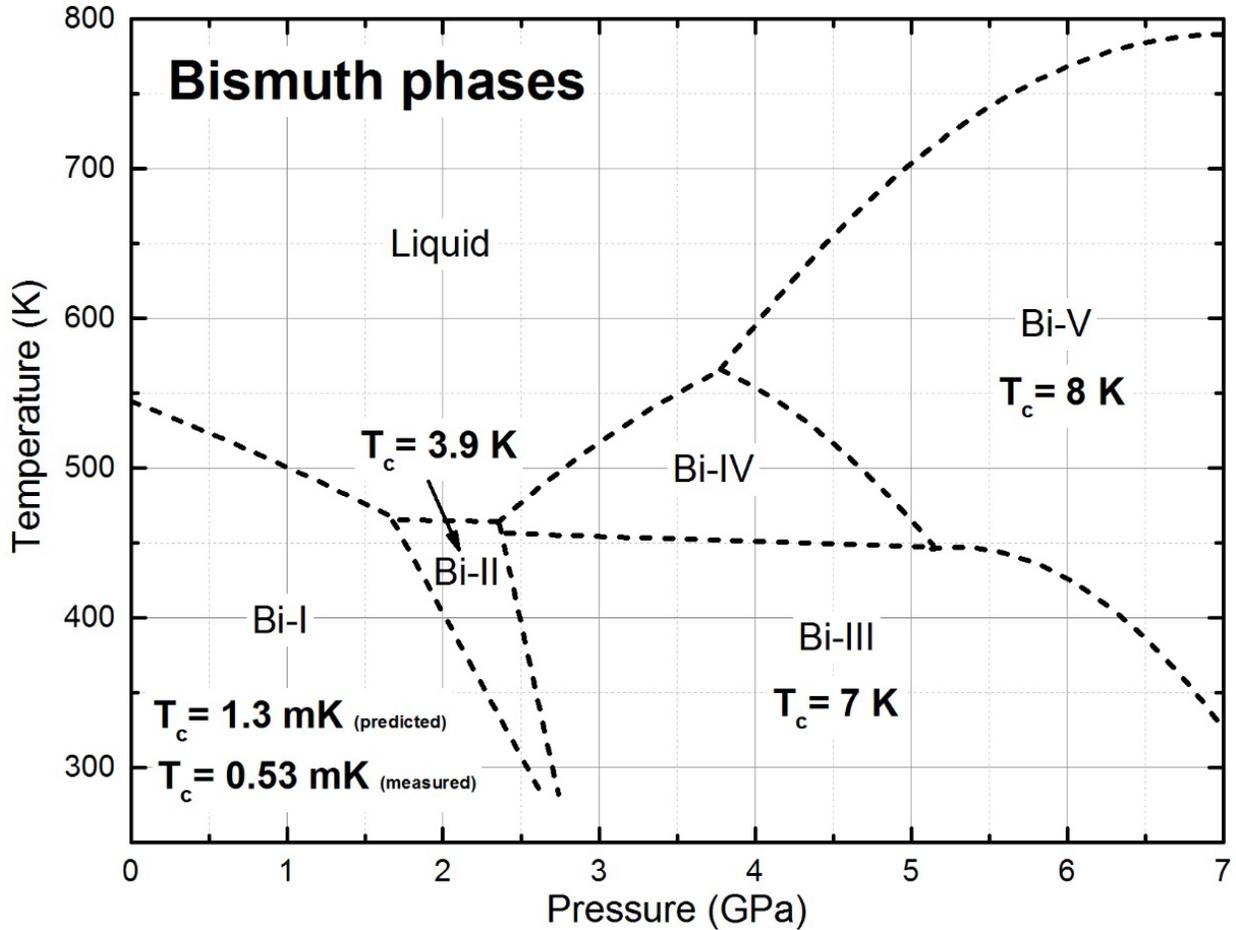

**Figure 1.** Phases of bismuth in the pressure-temperature plane taken from Ref. 7. The broken lines were obtained by linear and quadratic fits to experiment [8]. The experimental superconducting transition temperatures are included [9]. For the Wyckoff structure (Bi-I) [10] we report the $T_c$ we predicted [3] and the one subsequently measured [4].

The origin of these series of papers has to do with the question: why was amorphous Bi a superconductor and the crystalline Wyckoff phase [10] was not? No clue existed and to find an answer we undertook the computer generation of an amorphous sample of bismuth, whose pair distribution function agreed with experiment. We then calculated its $N(E)$ and its $F(\omega)$ and observed that the amorphous phase had a metallic-like character, since $N(E_F)$, the electronic density of states at the Fermi level, was reminiscent of a metallic system [3]. The amorphous phase had been found to be a superconductor with a $T_c \sim 6$ K so this led us to infer that the Wyckoff phase should become superconductive also, although at very low temperatures, prediction experimentally corroborated a year later [4].

If life can be simple, why complicate it? In this work we report a study of the electronic and vibrational properties of the solid phases of bismuth under pressure, calculated from first principles. We obtain $T_c$s very close to experiment, except for Bi-III, where the argued incommensurate nature of the crystal structure does not allow an unambiguous evaluation. For this phase we consider two different structures, one due to Chen *et al.* and the other constructed by us based on the results of McMahon *et al.* Superconductivity appears in most phases of bismuth because the Cooper pairing potential may not be drastically affected in going from one structure to the next as the pressure changes, and because the electronic properties, through the electron density of states *N(E)*, and the vibrational properties, through the vibrational density of states *F(ω)*, become the relevant factors. As Cooper points out in his decisive publication concerning the electron pairing as the phenomenon giving rise to superconductivity, "because of the similarity of the superconducting transition in a wide variety of complicated and differing metals, it is plausible to assume that the details of metal structure do not affect the qualitative features of the superconducting state" [11].

## 2. Procedure: the BCS approach

Calculating $T_c$ from basics has proven to be a very difficult task. The electron–phonon interactions are very complex processes and approximations have to be invoked that may mask the essential features of these interactions. However, it is possible to obtain results like the ones reported in this work that give an indication as to what materials may become superconductors if the *N(E)* and the *F(ω)* are known, and if the Cooper attraction sets in. In what follows we shall recur to some of our previous publications [3, 5-7] since the procedure is the same for these series of calculations.

We base our discussion on the BCS expression for the transition temperature:

$$T_c = 1.13\, \theta_D \exp[-1/N(E_F)V_0], \quad (1)$$

where $\theta_D$ is the Debye temperature and represents the role played by the vibrations, typified by *F(ω)*, the vibrational density of states (vDoS). $N(E_F)$ is the electron density of states (eDoS) at the Fermi level $E_F$, and $V_0$ is the Cooper pairing potential that binds pairs of electrons [2, 11]. Figure 2 shows $T_c$ as a function of the parameters that appear in Eqn. 1 for the specific value of the pairing potential calculated in this work, where the strong dependence of the transition temperature on the factor $N(E_F)$ can be observed.

This indicates that under certain circumstances $N(E_F)$ can play a more important role than $\theta_D$, especially when different phases of the same material are compared since in this situation it can be assumed that the strength of the pairing potential would not be altered much by the phase changes. Also, the Debye temperatures may not change drastically whereas the electronic properties could change radically, going from being a semimetal to becoming a conductor as is the case for bismuth [7].

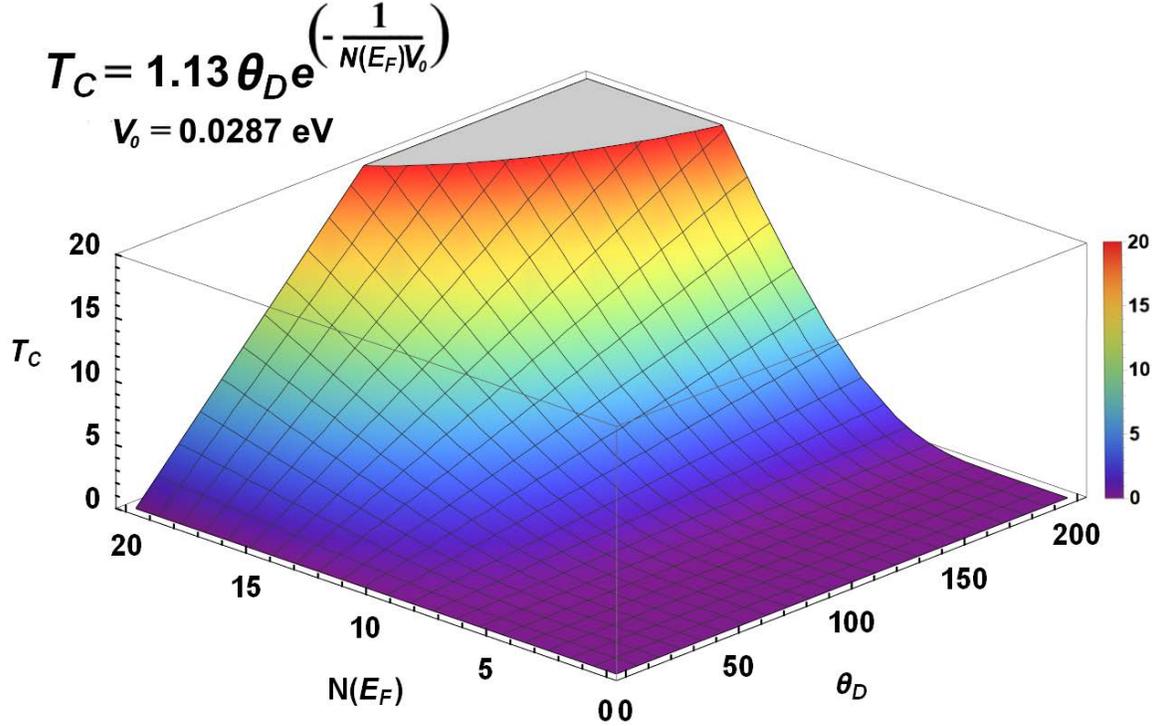

**Figure 2.** The superconducting transition temperature $T_c$ as a function of the electronic $N(E_F)$ and vibrational $\theta_D$ (the Debye temperature) parameters for the value of the Cooper pair potential $V_0$ calculated in this work.

From Eqn. 1, if the $N(E_F)$ and the $F(\omega)$ are known, together with the pairing potential $V_0$, the superconducting transition temperature can be obtained, as long as we can calculate $\theta_D$ from the function $F(\omega)$. The computational developments that have taken place in the recent decades allow the precise calculation of $N(E_F)$ and $F(\omega)$ for a variety of atomic structures of various materials; crystalline, amorphous, porous, liquids, etc. However, the precise calculation of the pairing potential $V_0$ has eluded the efforts of the simulational community.

Suppose that we know the transition temperature, the eDoS and the vDoS for a given phase of bismuth that we identify as $\Delta$ and want to obtain the transition temperature of a $\Gamma$ phase for which we also know the eDoS and the vDoS, but do not know the pairing potential for either one. Let us assume that the two phases of a given element have similar $V_0$s; we may consider for simplicity that the pairing potential is the same and this can be used to obtain the unknown $T_c$ of the phase under scrutiny. The facile procedure is as follows. All quantities related to the $\Gamma$ phase will be identified with a $\Gamma$ superscript in the BCS equation,

$$T_c^\Gamma = 1.13\, \theta_D^\Gamma\, \exp(-1/N(E_F)^\Gamma V_0).$$

Analogously, for the $\Delta$ phase we get

$$T_c^{\Delta} = 1.13 \, \theta_D^{\Delta} \, exp(-1/N(E_F)^{\Delta} V_0).$$

If we take the ratio $T_c^{\Gamma} / T_c^{\Delta}$ and find an expression for $T_c^{\Gamma}$ we obtain:

$$T_c^{\Gamma} = \{T_c^{\Delta}\}^{1/\eta} \, \delta \, \{1.13 \, \theta_D^{\Delta}\}^{(\eta-1)/\eta}, \tag{2}$$

where

$$N(E_F)^{\Gamma} = \eta \, N(E_F)^{\Delta} \quad \text{and} \quad \theta_D^{\Gamma} = \delta \, \theta_D^{\Delta}, \tag{3}$$

and the pairing potential $V_0$ cancelled out since we assumed it to be the same for all phases. How strong this supposition is will become apparent when we calculate the superconducting transition temperatures of all solid phases in what follows.

The Debye temperatures $\theta_D$, as required by Eqn. (1), is calculated from the vibrational spectra of each structure. For this we use an expression due to Grimvall, for the Debye frequency $\omega_D$ [12]

$$\omega_D = \exp\left[1/3 + \frac{\int_0^{\omega_{max}} \ln(\omega) F(\omega) \, d\omega}{\int_0^{\omega_{max}} F(\omega) \, d\omega}\right], \tag{4}$$

that we utilized in our previous work with good results [5-7]. $F(\omega)$ is the vDoS of the supercell under consideration, and $\omega_{max}$ is the maximum frequency of the corresponding vibrational spectrum. Using Eqn. 4, the calculations for $\theta_D = \hbar\omega_D / k_B$ give the values for each phase structure and with these results and those for $N(E_F)$ we are in a position to proceed. To obtain these results we removed the translational modes around 0 THz ($\omega \approx 0$) for all structures; this are more relevant the smaller the number of atoms in a supercell.

Now we may calculate the $T_c$ of the $\Gamma$ phase by using the values of the reference $\Delta$ phase. Since we have calculations of the eDoS and the vDoS for the amorphous bismuth structure, for the Wyckoff phase, for the bilayer structure and for the Bi-IV phase we can choose any of them as reference. To be consistent, we have been referring all previous calculations to the results obtained for the Wyckoff structure for which we obtained the $N(E_F)$, the $F(\omega)$, the $\theta_D$, and the $T_c$ by using the results for the amorphous structure that we studied originally. So from now on the Wyckoff phase shall be used as the $\Delta$ phase, *i.e.*, the reference phase, the phase we originally predicted to be superconductive; the $\Gamma$ phase shall be the corresponding solid phase under pressure whose $T_c$ we want to calculate.

### 3. The phases: calculations

#### 3.1 The Method

In what follows we shall report calculations obtained for the five phases of bismuth shown in Figure 1. For the sake of completeness, we include results previously published for the Bi-I and the Bi-IV phases to present an integrated work. Since the crystalline structures are experimentally known we shall use them to construct the corresponding crystalline supercells, with periodic boundary conditions, to proceed to the calculation of the quantities $N(E)$ and $F(\omega)$, from which $N(E_F)$ and $\theta_D$ shall be obtained. Once these quantities are known we use Eqn. (2) to calculate all the superconducting transition temperatures for the solid phases of bismuth under pressure, as a function of the reference structure $\Delta$.

As described elsewhere [7], the quantities $N(E)$ and $F(\omega)$ were obtained using the DMol3 code which is part of the Dassault Systèmes BIOVIA Materials Studio suite [13]. For the various supercells used, based on the experimental structures, a single-point energy calculation was performed first, using a double-numeric basis set and a fine mesh within the LDA-VWN approximation [14]; also, an unrestricted spin-polarized calculation for the energy was done. Since bismuth is a heavy element with many (83) electrons, the density-functional semi-core pseudo-potential (DSPP) approximation was used [15]. This pseudo-potential has been investigated by Delley where an all electron calculation is compared to the DSPP; the rms errors are essentially the same for both methods, 7.7 vs 7.5 [15]. Scalar relativistic corrections are incorporated in these pseudopotentials, essential for heavy atoms like Bi. Since DSPPs have been designed to generate accurate DMol3 calculations, it is expected that their use represents a good approximation; considerations of symmetry were left out for all structures. The calculation parameters were the same for the respective $N(E)$ and $F(\omega)$ for all structures, so meaningful comparisons can be made. For example, an energy convergence of $10^{-6}$ eV was used throughout, the real space cutoff was set to 6.0 Å, and the integration grid was set to fine; the calculations were carried out using a Monkhorst-Pack mesh adequate to each structure as shall be seen later for each phase. For the vibrational calculations, the finite-displacement approach was employed within DMol3, with a step size of 0.005 Å to calculate the Hessian using a finite-difference evaluation. Also, since the number of atoms in the supercells are different we have normalized our results to make them comparable and give $N(E)$ as the number of states per electron-volt per atom and $F(\omega)$ as the number of modes per frequency unit per atom everywhere.

For the eDoS the DMol3 analysis tools included in the Materials Studio suite were used, set to eV, and an integration method with a smearing width of 0.2 eV. The number of points per eV was 100. For the vDoS the results were analyzed with the OriginPro software, the calculated normal modes were imported in THz. To obtain the vDoS a frequency count with a 0.11 THz bin width was used, and the resulting bins were smoothed with a two-point FFT filter. The integral under the curve is normalized to three. The results for both densities of states are given per atom everywhere. Finally, the equations we shall use for the $\Gamma$ phase are

$$T_c^{\Gamma} = \{T_c^W\}^{1/\eta} \, \delta \, \{1.13 \, \theta_D^W\}^{(\eta-1)/\eta}, \qquad (5)$$

where we have explicitly included the Wyckoff phase as the reference phase. Also,

$$N(E_F)^{\Gamma} = \eta \, N(E_F)^W, \qquad (6)$$

and

$$\theta_D^{\Gamma} = \delta \, \theta_D^W. \qquad (7)$$

### 3.2 Antecedents

Solid Bi phases under pressure have been studied for decades [7]. Bismuth has been an appealing material, perhaps due to its low melting point and to its unusual physical properties. Since 1935 the phases of solid bismuth under pressure have been studied systematically. Bridgman, in his seminal work [16, 17], established the existence of several phases [18, 19] that have evolved into 5 presently accepted structures somewhat different to the original classification [20]. All 5 structures have been determined experimentally, inclusive the so-called incommensurate one of Bi-III. It is now taken for granted that these 5 phases extend to low temperatures, except for the ill-studied Bi-IV that exists in a very well-defined area in the P-T plane, 2.5 GPa ≤ P(Bi-IV) ≤ 5.0 GPa, and in the neighborhood of 500 K, [7].

Bismuth is a puzzling material; a versatile substance; it is the heaviest element of group 15, the highest atomic-number semimetal. At ambient pressure and temperature Bi is a crystalline solid, frustrated since it would like to be cubic but ends up being a rhombohedral (layered-like) structure, a semimetal for which the conducting properties are limited. At very low temperatures amorphous Bi (*a*-Bi) is stable and becomes a conductor, but more surprisingly it becomes a superconductor, and this was an incomprehensible result until we studied it and determined that the superconductivity is mainly due to changes in the electronic structure at the Fermi level [3]. It would seem that making it amorphous by lifting all the symmetry restrictions of the Wyckoff crystal frees the electrons and lets them Cooper-pair to superconduct.

In what follows we shall refer the reader to Ref. 20, Appendix A1, for the parameters that characterize the crystalline structures of the solid phases of bismuth under pressure used in this work, except for the proposed structures of Bi-III due to Chen *et al.* [21], and to McMahon *et al.* [22] since the debate exists as to what structure may be more representative of the phase. These parameters were determined at room temperature and we shall specify in each case what the pressure is for the parameters used so a precise structure is kept in mind. All pressure values have an uncertainty of 0.05 GPa and for *a*, *b* and *c* the uncertainties are less than 0.001 Å, and for *β* it is less than 0.01° [20]. Tables 1 and 2 list some crystallographic parameters taken from Ref. 20 for the phases, and some physical data calculated in this work.

**Table 1.** Crystallographic data for the 5 solid phases of bismuth. Bi-III is the so-called incommensurate structure and the debate still exists as to which commensurate one may be more representative. We consider both Chen *et al.* [21] and McMahon *et al.* [22] host(H)-guest(G) structures for this phase (For the 3:4 cell c = 12.49 Å).

| Phase | Unit Cell Parameters | | | | | | Space Group | |
|---|---|---|---|---|---|---|---|---|
| | $a$ [Å] | $b$ [Å] | $c$ [Å] | $\alpha$ [°] | $\beta$ [°] | $\gamma$ [°] | No. | Name |
| Bi-I | 4.55 | 4.55 | 11.86 | 90.00 | 90.00 | 120.00 | 166 | $R\bar{3}m$ |
| Bi-II | 6.65 | 6.09 | 3.29 | 90.00 | 110.37 | 90.00 | 12 | $C2/m$ |
| Bi-III (Chen *et al.*) | 8.66 | 8.66 | 4.24 | 90.00 | 90.00 | 90.00 | 85 | $P4/n$ |
| Bi-III (McMahon *et al.*) | 8.52 | 8.52 | 4.16(H) 3.18(G) | 90.00 | 90.00 | 90.00 | 1 | $P1$ |
| Bi-IV | 11.19 | 6.62 | 6.61 | 90.00 | 90.00 | 90.00 | 64 | $Cmca$ |
| Bi-V | 3.80 | 3.80 | 3.80 | 90.00 | 90.00 | 90.00 | 229 | $Im\bar{3}m$ |

**Table 2.** Physical data and supercell information for the 5 solid phases of bismuth. We consider both Chen *et al.* [21] and McMahon *et al.* [22] structures for the so-called incommensurate Bi-III phase. For Bi-IV we used two supercells, one (256 atoms) to calculate eDoS and the other (128) to obtain vDoS [7].

| Phase | Pressure [GPa] | Density [g/cm³] | Super Cell Multiplier | # Atoms | $N(E_F)$ [# e−states / eV × atom] | $\Theta_D$ [K] |
|---|---|---|---|---|---|---|
| Bi-I | 0.0 | 9.80 | 5x4x2 | 240 | 0.15 | 134.2 |
| Bi-II | 2.7 | 11.06 | 4x3x5 | 240 | 0.50 | 115.5 |
| Bi-III (Chen *et al.*) | 3.8 | 10.92 | 2x2x3 | 120 | 0.62 | 96.9 |
| Bi-III (McMahon *et al.*) | 4.2 | 12.25 | 2x2x1 | 128 | 0.45 | 144.4 |
| Bi-IV | 3.2 | 11.33 | 4(2)x2x2 | 256 (128) | 0.53 | 102.1 |
| Bi-V | 8.5 | 12.64 | 5x5x5 | 250 | 0.56 | 137.8 |

For the eDoS and the vDoS calculations we tried to construct supercells with the same number of atoms as closely as possible. The figures for the phases in this section have been drawn with a default bond length of 3.45 Å where the covalent radius is taken as 1.46 Å. For the Wyckoff phase and for Bi-IV we include our previously published results [3, 7] for the sake of completeness.

### 3.3 Bismuth I. The Wyckoff Phase

Both the amorphous bismuth structure and the crystalline one have been fully analyzed in Ref 3 and the crystalline one was revisited in Ref 7. The results are very similar in both references except for the fact that in the present work we have removed the translational modes. For the Wyckoff structure [10], Bi-I, we used a 6-atom crystalline cell that we multiply by 5x4x2 to obtain a supercell with 240 atoms for the calculations of $N(E)$ and of $F(\omega)$. The crystalline structure of this phase has two near-neighbor peaks located in 3.11 Å and 3.49 Å; these two peaks coalesce to give rise to a single wide peak for the amorphous sample [3]. The integration grid was set to fine and the calculations were carried out using a Monkhorst-Pack mesh of 2 × 3 × 3 in k-space. Since this phase is the reference phase, in this section we shall not reproduce $N(E)$ and $F(\omega)$ since they appear in the figures where comparisons are made (see also Refs. 3 and 7).

     As mentioned before, based on these ideas in Ref. 3 we carried out some estimations that led us to propose an upper bound for the superconducting transition temperature of the Wyckoff phase of Bi. We obtained a $T_c$ for crystalline Bi at atmospheric pressure of 1.3 mK, or lower, by comparing values of $N(E)$ and $F(\omega)$ between the Wyckoff phase and the amorphous phase, calculated using *ab initio* computational simulations. Several months later our predictions were corroborated experimentally ($T_c = 0.53$ mK*)* [4].

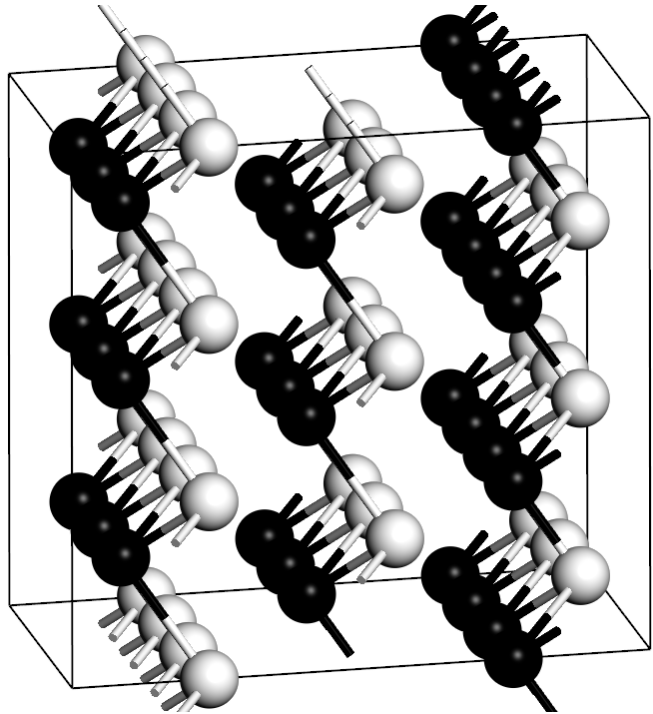

**Figure 3.** Representation of the rhombohedral structure of the Wyckoff phase displaying the bilayers. The bonds correspond to a default of 3.45 Å where the covalent radius is taken as 1.46 Å, as mentioned at the end of Section 3.2. This is the case for all structures.

For Bismuth I the pressure considered was 0 GPa, with 6 atoms in the hexagonal representation of the rhombohedral primitive cell. The following lattice parameters were used: $a = b = 4.55$ Å and $c = 11.86$ Å and with $\alpha = \beta = 90°$ and $\gamma = 120°$, and the space group is $R\bar{3}m$, as reported in Table 1. Figure 3 is a representation where the bilayers are clearly indicated.

### 3.4 Bismuth II. The non-low temperature Phase

Bismuth II is a phase that narrows as the temperature is decreased, Figure 1, and the fact that an experimental superconductive transition temperature has been reported, $T_c$ = 3.9 K, indicates that it remains a well-defined structure at very low temperatures [9]. The supercell used for the *ab initio* calculation of *N(E)* and *F(ω)* has 240 atoms, result of a 4x3x5 multiplication of the 4-atom unit cell and the parameters considered are $a$ = 6.65 Å, $b$ = 6.09 Å and $c$ = 3.29 Å, with $\alpha = 90°$, $\beta = 110.37°$ and $\gamma = 90°$ with a *C2/m* space group registered in Table 1, all at a pressure of 2.7 GPa [20]. A representation of the corresponding cell is shown in Figure 4 where the monoclinic structure can be observed, together with two (dark and light) distinct monoatomic layers displayed. The integration grid was set to fine and the calculations were carried out using a Monkhorst-Pack mesh of 2 × 3 × 3 in k-space.

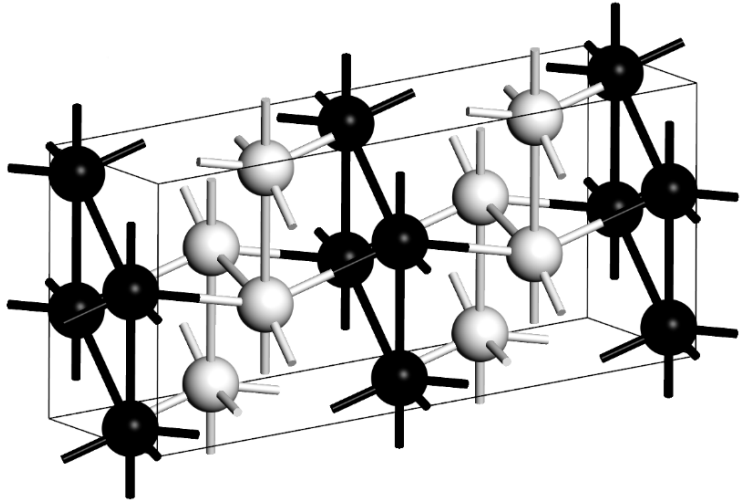

**Figure 4.** Representation of monoclinic *C2/m* structure of Bi-II where two distinct monoatomic layers (black and white spheres) can be observed.

The results obtained for the electronic density of states is given in Figure 5a where a comparison is made with the results obtained for the Wyckoff phase. The density of electron states at the Fermi level for the Wyckoff phase is 0.15 electrons per atom and for the Bi-II phase is 0.50 electrons per atom. The ratio $\eta$ is 3.33. In Figure 5b a comparison of the results obtained for the vibrational density of states is presented. The calculations for the Debye temperatures give 134.2 K for the Wyckoff phase and 115.5 K for Bi-II. The ratio $\delta$ is 0.87. The calculated transition temperature for the Bi-II phase is 3.6 K which is to be compared with the experimental result of 3.9 K.

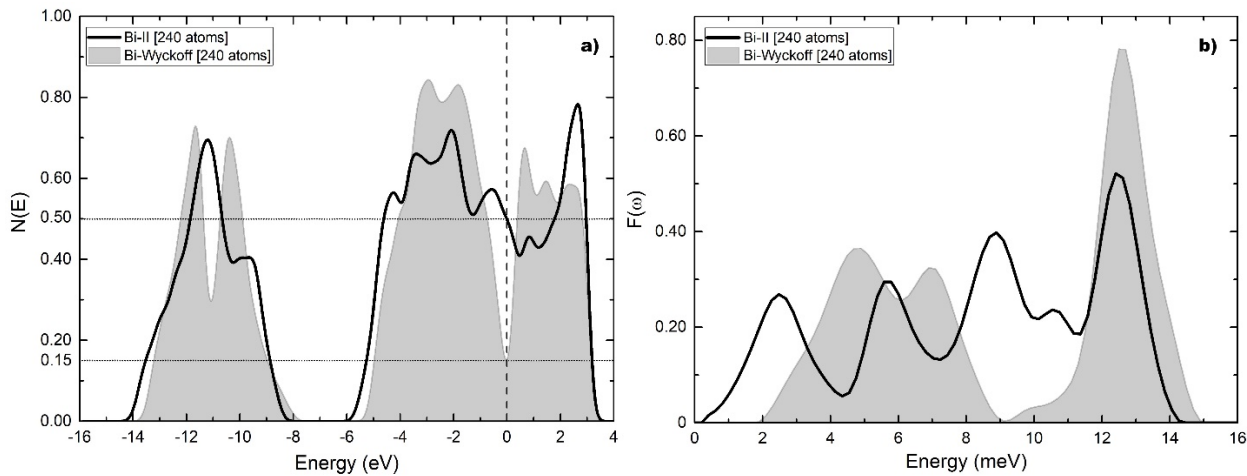

**Figure 5.** a) $N(E)$ and b) $F(\omega)$ for Bi-II (black solid line) compared to the results for the Wyckoff structure (grey). Since the supercells may have a different number of atoms we plot the results per atom everywhere to compare them meaningfully.

### 3.5 Bismuth III. The incommensurate Phase

Paraphrasing Hamlet: **To commensurate, or not to commensurate, that is the question.**

Bismuth III is a low temperature phase that extends to pressures higher than 7 GPa, Figure 1. The experimental superconductive transition temperature reported, $T_c$ = 7 K, is the highest second to that of Bi-V. This phase has been somewhat controversial since it was described as a regular periodic cell when first identified [21], and nowadays it is considered as an incommensurate structure [20, 22]. Since an incommensurate structure is very difficult to represent and even more difficult to calculate, there have been attempts to propose representative commensurate periodic crystalline structures that supposedly would reflect with some precision the real structure [23]. This gave rise to two structures, one due to Chen and coworkers [21], the other due to McMahon and coworkers [22]; Häussermann *et al.* proposed a commensurate approximation [24]. In this work we consider Chen´s representation and, based on the 3:4 Häussermann approximation, we construct a commensurate one for McMahon´s data and calculate their properties and their superconducting transition temperature. As we shall see, the structure proposed by Chen *et al.* leads to a $T_c$ closer to experiment than ours.

The Chen supercell used for the *ab initio* calculation of $N(E)$ and $F(\omega)$ has 120 atoms, result of a 2x2x3 multiplication of the 10-atom representative cell. Figure 6 is a representation where the tetragonal structure is illustrated. The parameters, registered in Table 1, at a pressure of 3.8 GPa, are: $a = b = 8.66$ Å, $c = 4.24$ Å; $\alpha = \beta = \gamma = 90°$ and the space group is *P4/n*, [21]. The integration grid was set to fine and the calculations were carried out using a Monkhorst-Pack mesh of 2 × 3 × 3 in k-space.

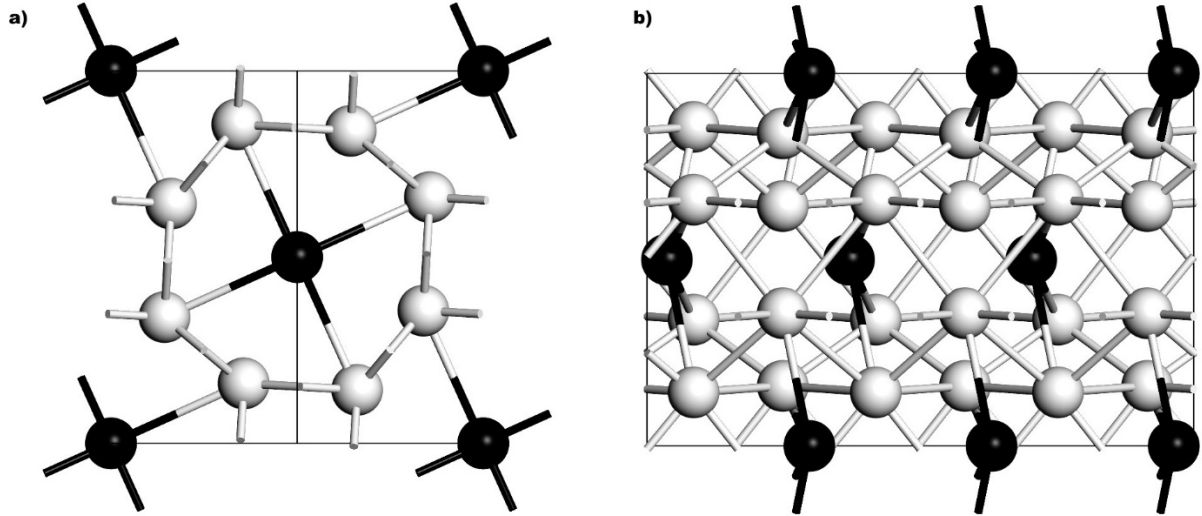

**Figure 6.** Tetragonal structure proposed by Chen *et al.* [21] to describe the experimentally studied incommensurate Bi-III phase at 3.8 GPa. a) The x-y plane of the structure b) The horizontal in-the-plane axis is the z-axis of the structure.

On the other hand, we propose a slightly different commensurate host-guest structure, after McMahon *et al.* concluded that Bi III is, in fact, formed by tetragonal host-guest structures incommensurate with one another along the z-axis [22]. The supercell we used for the *ab initio* calculation of *N(E)* and *F(ω)* has 128 atoms, result of a 2x2x1 multiplication of our 32-atom representative cell and the parameters are: $a = b = 8.52$ Å, $c_H = 4.2$ Å and $c_G = 3.1$ Å for the host and guest structures, respectively; $α = β = γ = 90°$ and the space group is *P1,* at a pressure of 4.2 GPa (compare with Table 1) [22]. The integration grid was set to fine and the calculations were carried out using a Monkhorst-Pack mesh of 2 × 3 × 3 in k-space. Our basic 32-atom cell is represented in Figure 7. The host structure is tetragonal with 8 atoms in the unit cell (white spheres), whereas the guest structure is body centered tetragonal with two atoms in the unit cell (black spheres).

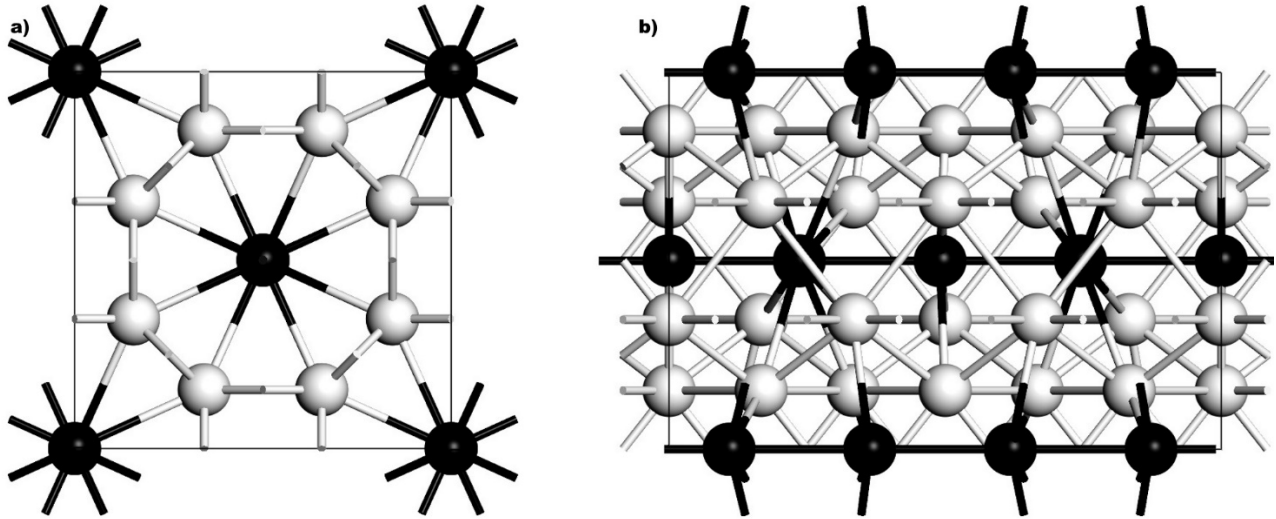

**Figure 7.** Our commensurate Bi-III 3:4 host-guest structure based on McMahon incommensurate structure [22], at 4.2 GPa. a) The x-y plane of the structure b) The horizontal in-the-plane axis is the z-axis of the structure.

For the Chen structure the eDoS is given in Figure 8(a) compared to the Wyckoff phase. The eDoS for the Wyckoff phase is 0.15 electrons per atom and the one for the Bi-III phase is 0.62 electrons per atom. The ratio $\eta$ is 4.13. In Figure 8(b) the vDoS are given. The Debye temperatures are 134.2 K for the Wyckoff phase as before, and 96.4 K for Bi-III. The ratio $\delta$ is 0.72. The calculated transition temperature for this structure is 6.5 K which is to be compared with the experimental result of 7 K.

Figures 9 are the corresponding results for our structure, *a la* Häussermann, based on the data reported by McMahon *et al.*, compared to Wyckoff´s. The density of electron states for this structure at the Fermi level is 0.45 electrons per atom and the ratio $\eta$ is 3.00. The Debye temperature 144.4 K and the ratio $\delta$ is 1.08. The calculated transition temperature is then 3.5 K.

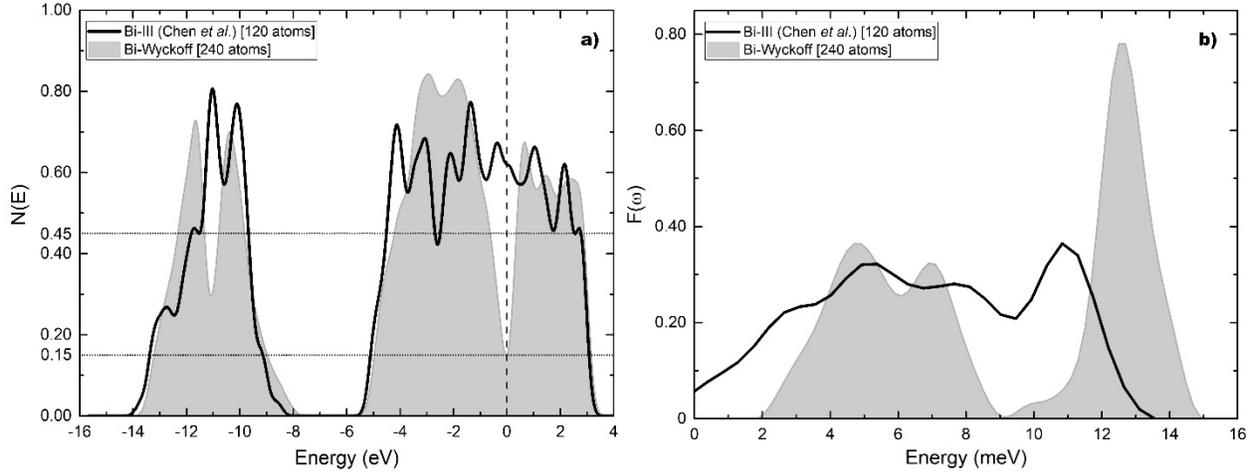

**Figure 8.** a) *N(E)* and b) *F(ω)* for Bi-III (black solid line) according to Chen *et al.*, compared to the results for the Wyckoff structure (grey). Negative energy vibrational modes were removed from Bi-III.

So, what then? **to commensurate or not to commensurate**? Doing the incommensurate cell would be more representative but given the difficulty [23] we decided to calculate Bi-III *a la* Chen that leads to $T_c$ of 6.5 K compared to the experimental result of 7 K, as opposed to a $T_c$ of 3.5 K for our commensurate cell *a la* Häussermann. Chen´s seems more adequate for our purposes but it is necessary to find a better commensurate representation of McMahon´s results

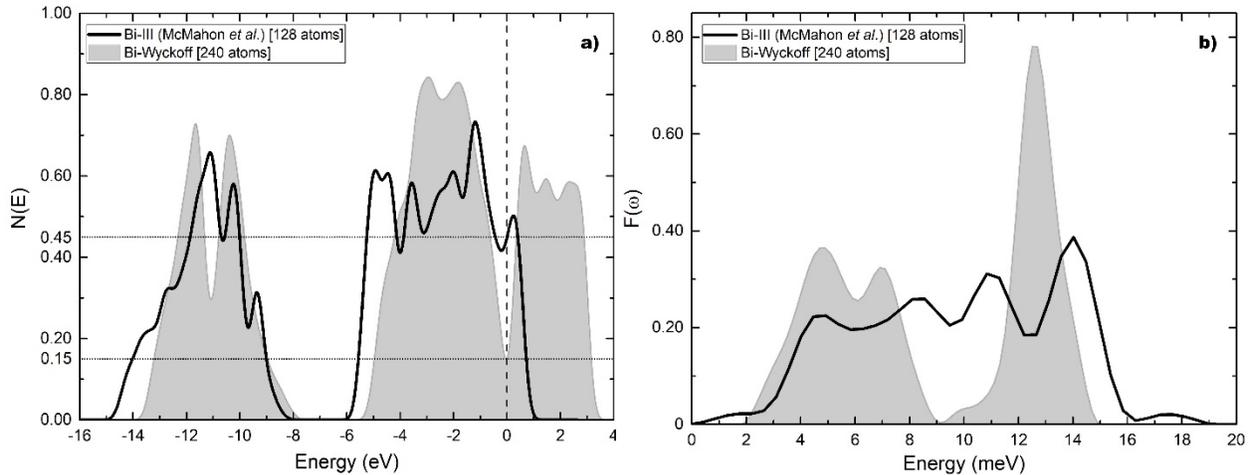

**Figure 9.** a) *N(E)* and b) *F(ω)* for Bi-III (black solid line) according to our commensurate representation of McMahon´s results, compared to the Wyckoff structure (grey).

### 3.6 Bismuth IV. The "non-superconducting" phase

Here we reproduce some results reported in Ref. 7, for completeness. Experimentally, Bismuth IV has not been found to be a superconductor. However, since all other phases are superconductive, it seems highly probable that Bi-IV would also be, provided the proper conditions are obtained. That is why we decided to investigate this phase using the same facile approach that we describe in this work. We found that, in fact, this phase could become a superconductor if its crystalline structure could be maintained while cooling it to a transition temperature of 4.25 K. Corroboration is pending.

Experimental results by Chaimayo *et al.* [25] for bismuth IV have confirmed that its crystalline structure is orthorhombic at 465 K and 3.2 GPa with the space group given as *Cmca*, with $a$ = 11.19 Å, $b$ = 6.62 Å and $c$ = 6.61 Å, and with the crystalline angles given by $α = β = γ = 90°$, see Table 1. A 2 x 2 x 3 Monkhorst-Pack scheme in k-space was used and a fine integration grid utilized. The 16-atom crystalline structure of this phase is represented in Figure 10 with the bilayers depicted with white spheres and intercalated between gray and black layers. By multiplying this cell 4 x 2 x 2 times a supercell with 256 atoms is obtained. It is this supercell that was used to calculate *N(E)*. To obtain *F(ω)* a supercell with 128 atoms was used, result of a 2 x 2 x 2 multiplication of the depicted cell [7]. In 1958 Bundy identified this phase and it was then assumed to be cubic [19].

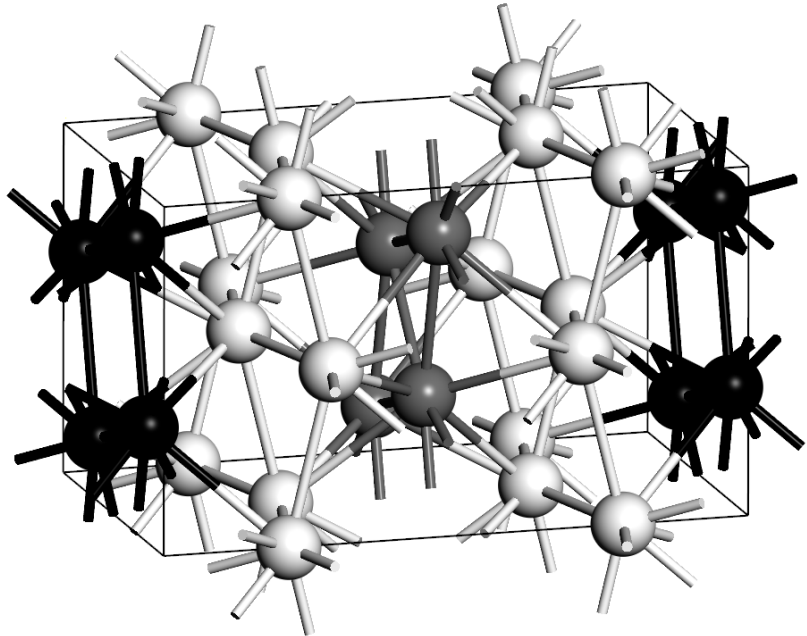

**Figure 10.** Crystalline cell of Bi-IV with 16 atoms, taken from Ref. 7. The bilayers (white spheres) and the layers with grey and black spheres are shown.

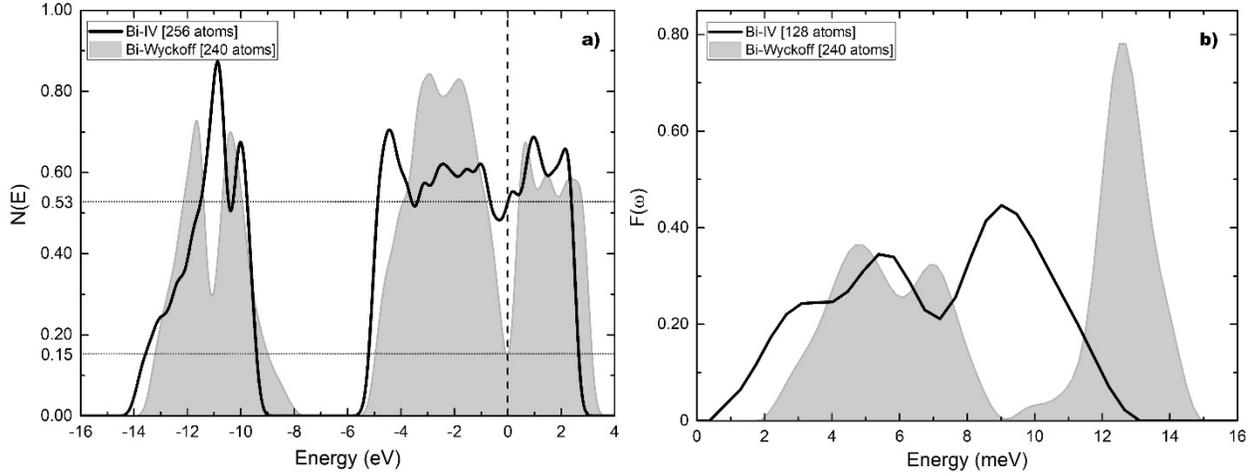

**Figure 11.** a) N(E) and b) F(ω) for Bi-IV (black solid line) compared to the Wyckoff structure (grey) as previously reported in Ref. 7.

The results obtained for the eDoS and the vDoS of the Chaimayo structure are given in Figures 11. Figure 11(a) shows that *N(E)* at the Fermi level is, as always, 0.15 electrons per atom for the Wyckoff phase and 0.53 electrons per atom for Bi-IV; the ratio $\eta$ being 3.53. Figure 11(b) presents a comparison of the two vibrational densities of states. The calculations for the Debye temperatures lead to 134.2 K for the Wyckoff phase and 102.1 K for Bi-IV; the ratio $\delta$ is 0.76. The predicted superconducting transition temperature for this crystalline phase is 4.25 K, result to be corroborated.

The $\theta_D$ now obtained for the Wyckoff phase, 134.2 K, is somewhat larger than that reported in Ref. 3, 129 K, since as mentioned before we have ignored the translational "static" modes ($\omega \approx 0$), proportionally more relevant as the number of atoms in the supercell diminishes. As mentioned before, the experimental values for $\theta_D$ reported by DeSorbo [26] for crystalline bismuth at ambient pressure goes from 140 K at high temperatures to 120 K at low temperatures. Our calculations indicate that $\theta_D$ = 134.2 K, a value that lies between the experimental ones. It should be noticed that the vibrational spectrum for Bi-IV is more localized than that for the Wyckoff phase, although one would expect it to be more extended since the application of pressure would bring the atoms together (the mass density does increase, 9.80 *vs* 11.33 g/cm$^3$, Table 2) augmenting the force constant; evidently however, the change in the crystalline structure supersedes this factor.

### 3.7 Bismuth V. The high pressure phase

The structure of Bi-V at a pressure of 8.5 GPa is given by $a = b = c = 3.80$ Å, and the angles are $\alpha = \beta = \gamma = 90°$ and, as indicated in Figure 12, the structure is a body-centered cubic one (white spheres) with the space group *Im$\bar{3}$m*. The supercells used for the calculation of *N(E)* and *F(ω)* have 250 atoms, result of a 5x5x5 multiplication of the 2-atom unit cell. A Monkhorst-Pack mesh of 3 × 3 × 3 in k-space was used.

The results for *N(E)* and *F(ω)* for both the Wyckoff and Bi-V phases are presented in Figures 13 (a) and (b), respectively. The calculations give a result of 0.56 electrons per atom at the Fermi level, which when compared to the value for the Wyckoff phase (0.15), Figure 13(a), leads to a ratio $\eta$ of 3.73. In Figure 13(b) a comparison of the results obtained for the vibrational density of states is given. As before the Debye temperature is 134.2 K for the Wyckoff phase; the Debye temperature for Bi-V is 137.8 K and the ratio $\delta$ is 1.03. The calculated superconducting transition temperature for this phase is 6.8 K compared with the experimental result of 8 K. The vibrational spectrum for Bi-V has practically lost the gaps or pseudogaps observed in *F(ω)* for other phases, since the bcc crystal lacks the layered arrangements of other structures.

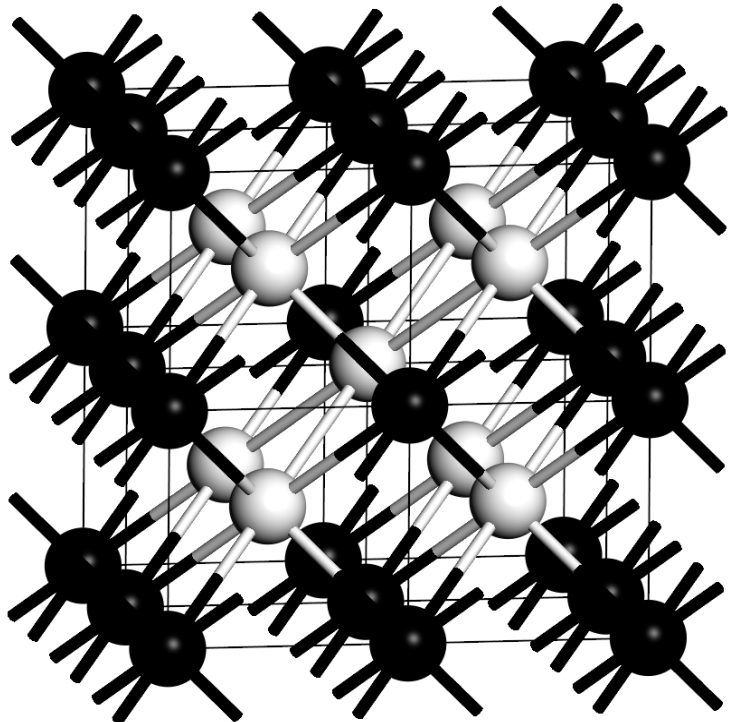

**Figure 12.** The body centered cubic structure of the Bi V phase can be seen in white [20].

## 4. Results and Discussion

Figure 14 shows the diagram of the bismuth phases under pressure together with the experimentally determined transition temperatures and our predictions and calculations. The agreement is satisfactory and indicates several features. The fact that the calculations were conducted under the BCS approximation indicates that this approach is adequate for these materials. It has been argued that bismuth is a strong coupling substance and that, in fact, McMillan equation [27] should be used when dealing with this material. Tunneling experiments conducted more than 50 years ago by Smith and coworkers [28] showed that for amorphous bismuth, $2\Delta_0/k_BT_c =$ 4.60 indicating a strong coupling regime, compared to the limit of weak coupling $2\Delta_0/k_BT_c =$ 3.5; however, the 5 bismuth phases under pressure, including Bi-I may not be strong coupling since our calculations agree well with experiment and we hope that our predictions will be corroborated by experiment, as was the case for the Wyckoff structure. Table 3 is a compendium of parameters for these solid phase structures.

Also, the comparison of the results for the higher-pressure phases was made against the Wyckoff phase. Clearly this comparison could have been made against other structure, based on experiment, or against the amorphous, etc. However, being systematic gives consistency.

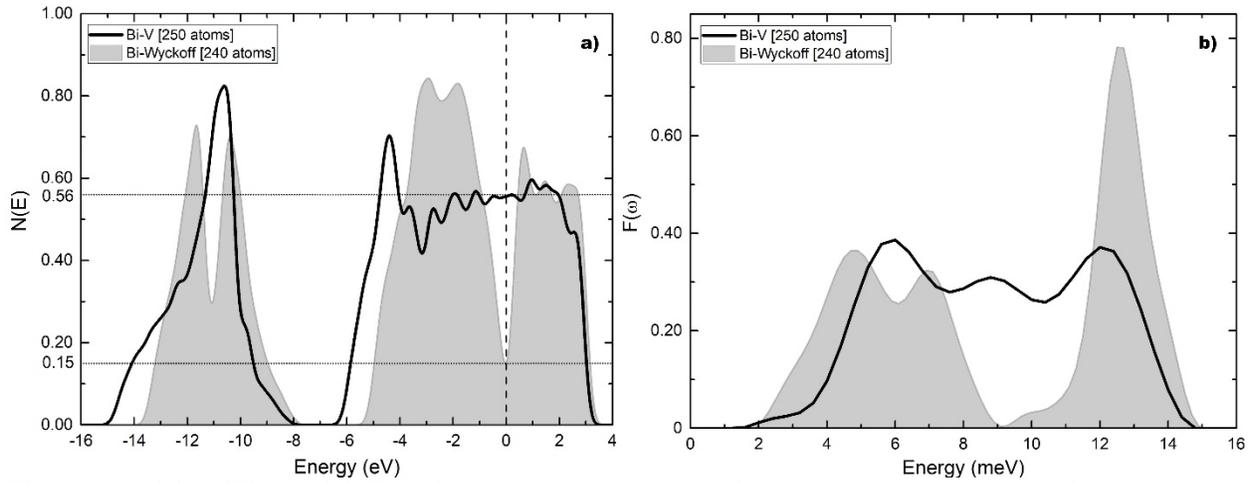

**Figure 13.** (a) *N(E)* and (b) *F(ω)* for Bi-V (black solid line) compared to the results for the Wyckoff structure (grey).

All our calculations are based on a reasonable assumption, the fact that the Cooper pairing potential is assumed to be the same for all phases. Since there are no estimates for the value of this potential for bismuth we decided to obtain it from the BCS equation and from the calculated results for *N(E_F)* and *F(ω)* (or *θ_D*) and the superconducting transition temperatures reported in Table 3, for each phase. As expected $V_o$ should be a constant and the same for all phases (as was assumed). In Figure 15 we plot this value for the five phases, acknowledging that for Bi-III two structures were considered. The calculation is carried out with both, the experimental transition temperatures and the superconducting transition temperatures obtained by us. It can be observed that the average of $V_o$ excluding both our representation of Bi-III and the Wyckoff phase, gives 0.0287 eV, a value that we used in the graph of Figure 2. The result for our representation of the Bi-III commensurate structure gives results away from experiment so this structure should be analyzed with more care. However, the result for the Wyckoff structure is considerably off. We have two possible explanations; the first deals with the small value found for *N(E_F)* for this phase so small variations of this quantity are proportionally very important, enough to make $V_o$ move from 0.0287 to close to 0.32. The second explanation is related to the possibility that this phase is strong coupling and that therefore is not adequately described by BCS. More work should be done to investigate this point and its implication on the $T_C$ predicted for this Bi-I phase.

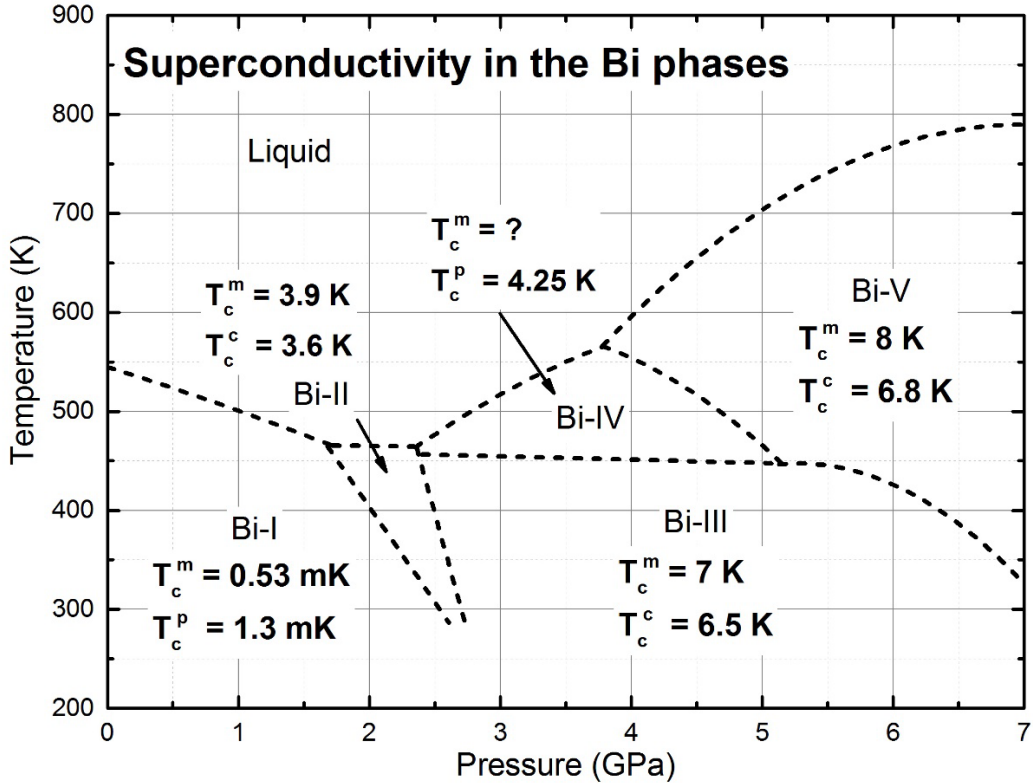

**Figure 14.** Bismuth phases under pressure with predicted $T_c^p$, calculated $T_c^c$, and experimentally determined $T_c^m$, superconducting transition temperatures. The Bi-III structure used was the one reported by Chen *et al.* [21].

**Table 3.** Values for the parameters $\eta$ and $\delta$ defined in Equations (6) and (7) for the various phases (See Table 2 also). The superconducting transition temperatures are also included. The temperatures predicted $T_c^{\,p}$ are explicitly given.

| Phase | $\eta$ | $\delta$ | $T_c^{\,p}$ [K] | $T_c^{\,c}$ [K] | $T_c^{\,m}$ [K] |
|---|---|---|---|---|---|
| Bi I | 1.00 | 1.00 | 0.0013 | - | 0.00053 |
| Bi II | 3.33 | 0.87 | - | 3.6 | 3.9 |
| Bi III (Chen *et al.*) | 4.13 | 0.72 | - | 6.5 | 7 |
| Bi III (Our structure) | 3.00 | 1.08 | - | 3.5 | 7 |
| Bi IV | 3.53 | 0.77 | 4.25 | - | - |
| Bi V | 3.73 | 1.03 | - | 6.8 | 8 |

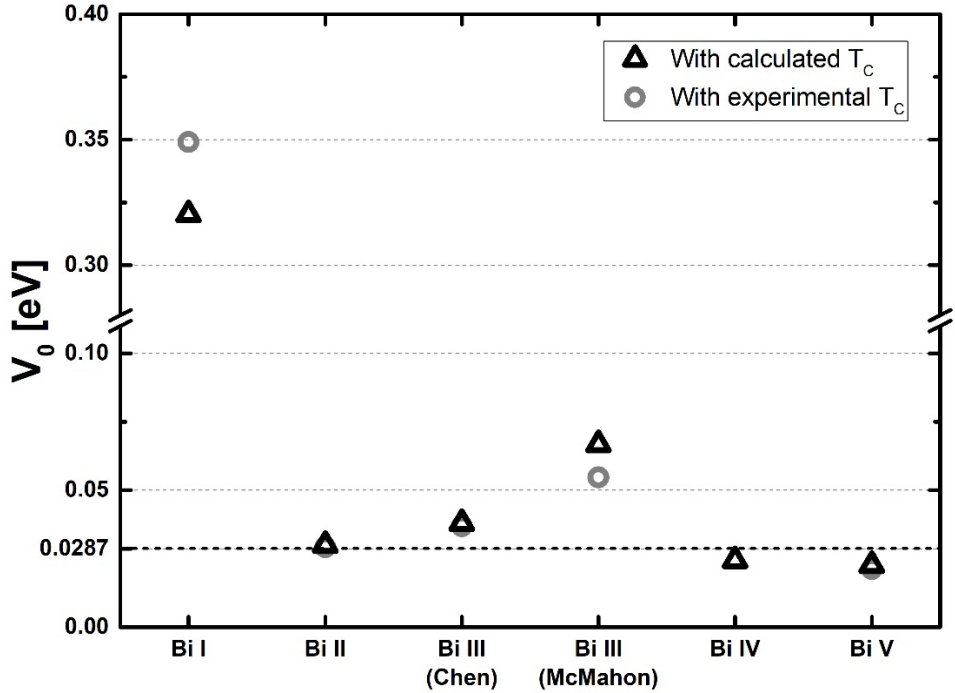

**Figure 15.** Cooper pairing potentials, $V_o$, for the solid phases of Bi under pressure, obtained from the product $N(E_F)\,V_o$.

## 5. Conclusions

Two important assumptions have been made when predicting the superconducting transition temperatures of the solid phases of Bismuth under pressure. These are:

*Assumption 1.-* All the phases considered can be described by the Bardeen, Cooper, Schrieffer equation for the superconducting transition temperature. This implies that we consider their superconductivity as phononic, or the conventional type;

*Assumption 2.-* Moreover, we assume that the Cooper pairing potential $V_0$ is the same for all phases, since the material is the same albeit in different crystalline structures and for different densities.

      However, the fact that we have been able to predict the superconducting transition temperature of Bi-I, already corroborated by experiment, and the fact that we have adventured predictions for the superconductivity in bismuth bilayers and for the phase Bi-IV, not experimentally studied yet, gives us confidence that, at least for bismuth, our approach is sound. The present work ascertains the validity of our approach, since the calculated superconducting transition temperatures are all very close to experiment. On the other hand, the assumption that the pairing potential is the same for all phases *a priori*, is validated by the calculation of its value *a posteriori*. The average of $V_0$ is 0.029 eV and when considering the factor ($N(E_F)$ $V_0$) this is much smaller than 1 for all phases, a condition indicating that we are in the weak coupling limit. The pairing mechanism considered is due to phonons as indicated by the factor $\theta_D$ that appears in the BCS expression, Eqn. 1, and in Figure 2.

      Calculating the pairing potential is a challenge. Computational simulations are helping to accomplish this but meanwhile, using the present approach, we can obtain reasonable approximations to the superconducting transition temperatures if one deals with the same material in different phases, case at point. Also, computational simulations have evolved favorably to allow one to obtain reliable results for the electronic and vibrational densities of states to be able to compare them meaningfully and infer conclusions. As far as the vibrational properties are concerned, one can observe interesting changes in $F(\omega)$ for all phases associated to the presence of low dimensional structures and, in some of them, associated to the presence of low frequency modes, often invoked as very important in the manifestation of superconducting properties.

      Certainly, the electronic and vibrational properties are conducive to propitiating superconductivity if the pairing potential manifests itself and for that the temperature has to be low enough until superconductivity appears; for most of them the liquid helium temperature is adequate, as shown in Figure 14. The results obtained and displayed in this figure, gives us confidence in our procedure. Other materials should be considered to see the breadth of our surmises.

**Data availability:** The datasets generated and analysed during the current study are available from the corresponding on reasonable request.

**Acknowledgements**
I.R. and D.H.R. acknowledge CONACyT for supporting their graduate studies. A.A.V., R.M.V. and A.V. thank DGAPA-UNAM for continued financial support to carry out research projects IN101798, IN100500, IN119908, IN112211, IN110914 and IN104617. M.T. Vázquez and O. Jiménez provided the information requested. Simulations were partially carried out in the Supercomputing Center of DGTIC-UNAM.

**Author Contributions**
A.A. Valladares, A. Valladares and R.M. Valladares conceived this research and designed it with the participation of I. Rodriguez and D. Hinojosa-Romero. I. Rodriguez and D. Hinojosa-Romero constructed the superlattices and calculated the electronic and vibrational densities of states, the superconducting transition temperatures and the Cooper pairing potential. All authors discussed and analyzed the results. A.A. Valladares wrote the first draft and the other authors enriched the manuscript.

**Additional Information**
**Competing Interests:** The authors declare no competing interests.

**Figure Legends**

**Figure 1.** Phases of bismuth in the pressure-temperature plane taken from Ref. 7. The broken lines were obtained by linear and quadratic fits to experiment [8]. The experimental superconducting transition temperatures are included [9]. For the Wyckoff structure (Bi-I) [10] we report the Tc we predicted [3] and the one subsequently measured [4].

**Figure 2.** The superconducting transition temperature $T_c$ as a function of the electronic $N(E_F)$ and vibrational $θ_D$ (the Debye temperature) parameters for the value of the Cooper pair potential $V_0$ calculated in this work.

**Figure 3.** Representation of the rhombohedral structure of the Wyckoff phase displaying the bilayers. The bonds correspond to a default of 3.45 Å where the covalent radius is taken as 1.46 Å, as mentioned at the end of Section 3.2. This is the case for all structures.

**Figure 4.** Representation of monoclinic *C2/m* structure of Bi-II where two distinct monoatomic layers (black and white spheres) can be observed.

**Figure 5.** a) *N(E)* and b) *F(ω)* for Bi-II (black solid line) compared to the results for the Wyckoff structure (grey). Since the supercells may have a different number of atoms we plot the results per atom everywhere to compare them meaningfully.

**Figure 6.** Tetragonal structure proposed by Chen *et al.* [21] to describe the experimentally studied incommensurate Bi-III phase at 3.8 GPa. a) The x-y plane of the structure b) The horizontal in-the-plane axis is the z-axis of the structure.

**Figure 7.** Our commensurate Bi-III 3:4 host-guest structure based on McMahon incommensurate structure [22], at 4.2 GPa. a) The x-y plane of the structure b) The horizontal in-the-plane axis is the z-axis of the structure.

**Figure 8.** a) *N(E)* and b) *F(ω)* for Bi-III (black solid line) according to Chen *et al.*, compared to the results for the Wyckoff structure (grey). Negative energy vibrational modes were removed from Bi-III.

**Figure 9.** a) *N(E)* and b) *F(ω)* for Bi-III (black solid line) according to our commensurate representation of McMahon´s results, compared to the Wyckoff structure (grey).

**Figure 10.** Crystalline cell of Bi-IV with 16 atoms, taken from Ref. 7. The bilayers (white spheres) and the layers with grey and black spheres are shown.

**Figure 11.** a) *N(E)* and b) *F(ω)* for Bi-IV (black solid line) compared to the Wyckoff structure (grey) as previously reported in Ref. 7.

**Figure 12.** The body centered cubic structure of the Bi V phase can be seen in white [20].

**Figure 13.** a) *N(E)* and b) *F(ω)* for Bi-V (black solid line) compared to the results for the Wyckoff structure (grey).

**Figure 14.** Bismuth phases under pressure with predicted $T_c{}^p$, calculated $T_c{}^c$, and experimentally determined $T_c{}^m$, superconducting transition temperatures. The Bi-III structure used was the one reported by Chen *et al.* [21].

**Figure 15.** Cooper pairing potentials, $V_o$, for the solid phases of Bi under pressure, obtained from the product $N(E_F) V_o$.

**Tables**

**Table 1.** Crystallographic data for the 5 solid phases of bismuth. Bi-III is the so-called incommensurate structure and the debate still exists as to which commensurate one may be more representative. We consider both Chen *et al.* [21] and McMahon *et al.* [22] host(H)-guest(G) structures for this phase (For the 3:4 cell c = 12.49 Å).

| Phase | Unit Cell Parameters | | | | | | Space Group | |
|---|---|---|---|---|---|---|---|---|
| | $a$ [Å] | $b$ [Å] | $c$ [Å] | $\alpha$ [°] | $\beta$ [°] | $\gamma$ [°] | No. | Name |
| Bi-I | 4.55 | 4.55 | 11.86 | 90.00 | 90.00 | 120.00 | 166 | $R\bar{3}m$ |
| Bi-II | 6.65 | 6.09 | 3.29 | 90.00 | 110.37 | 90.00 | 12 | $C2/m$ |
| Bi-III (Chen *et al.*) | 8.66 | 8.66 | 4.24 | 90.00 | 90.00 | 90.00 | 85 | $P4/n$ |
| Bi-III (McMahon *et al.*) | 8.52 | 8.52 | 4.16(H) 3.18(G) | 90.00 | 90.00 | 90.00 | 1 | $P1$ |
| Bi-IV | 11.19 | 6.62 | 6.61 | 90.00 | 90.00 | 90.00 | 64 | $Cmca$ |
| Bi-V | 3.80 | 3.80 | 3.80 | 90.00 | 90.00 | 90.00 | 229 | $Im\bar{3}m$ |

**Table 2.** Physical data and supercell information for the 5 solid phases of bismuth. We consider both Chen *et al.* [21] and McMahon *et al.* [22] structures for the so-called incommensurate Bi-III phase. For Bi-IV we used two supercells, one (256 atoms) to calculate eDoS and the other (128) to obtain vDoS [7].

| Phase | Pressure [GPa] | Density [g/cm³] | Super Cell Multiplier | # Atoms | $N(E_F)$ $\left[\frac{\# e - states}{eV \times atom}\right]$ | $\Theta_D$ [K] |
|---|---|---|---|---|---|---|
| Bi-I | 0.0 | 9.80 | 5x4x2 | 240 | 0.15 | 134.2 |
| Bi-II | 2.7 | 11.06 | 4x3x5 | 240 | 0.50 | 115.5 |
| Bi-III (Chen *et al.*) | 3.8 | 10.92 | 2x2x3 | 120 | 0.62 | 96.9 |
| Bi-III (McMahon *et al.*) | 4.2 | 12.25 | 2x2x1 | 128 | 0.45 | 144.4 |
| Bi-IV | 3.2 | 11.33 | 4(2)x2x2 | 256 (128) | 0.53 | 102.1 |
| Bi-V | 8.5 | 12.64 | 5x5x5 | 250 | 0.56 | 137.8 |

**Table 3.** Values for the parameters $\eta$ and $\delta$ defined in Equations (6) and (7) for the various phases (See Table 2 also). The superconducting transition temperatures are also included. The temperatures predicted $T_c^p$ are explicitly given.

| Phase | $\eta$ | $\delta$ | $T_c^p$ [K] | $T_c^c$ [K] | $T_c^m$ [K] |
|---|---|---|---|---|---|
| Bi I | 1.00 | 1.00 | 0.0013 | - | 0.00053 |
| Bi II | 3.33 | 0.87 | - | 3.6 | 3.9 |
| Bi III (Chen *et al.*) | 4.13 | 0.72 | - | 6.5 | 7 |
| Bi III (Our structure) | 3.00 | 1.08 | - | 3.5 | 7 |
| Bi IV | 3.53 | 0.77 | 4.25 | - | - |
| Bi V | 3.73 | 1.03 | - | 6.8 | 8 |